\documentclass[
twocolumn,
showpacs,preprintnumbers,
amsmath,amssymb,
aps,
pra,
 ]{revtex4-1}

\newcommand{\un}[2]{#1\,\mathrm{#2}}
\usepackage[T1]{fontenc}
\usepackage{graphicx}
\usepackage{dcolumn}
\usepackage{bm}
\usepackage{geometry}
\usepackage{color}

\begin{document}

\title{Energy levels of Th$^+$ between 7.3 and 8.3\,eV}

\author{O. A. Herrera-Sancho, N. Nemitz, M. V. Okhapkin, and E. Peik}

 \email{Corresponding author E-mail address: ekkehard.peik@ptb.de}
 \affiliation{Physikalisch-Technische Bundesanstalt, Bundesallee 100, 38116 Braunschweig, Germany}

\date{\today}

\begin{abstract}
Using resonant two-step laser excitation of trapped $^{232}$Th$^+$ ions, we observe 43 previously unknown energy levels within the energy range from $7.3$ to $8.3$~eV. The high density of states promises a strongly enhanced electronic bridge excitation of the $^{229m}$Th nuclear state that is expected in this energy range.
From the observation of resonantly enhanced three-photon ionization of Th$^+$, the second ionization potential of thorium can be inferred to lie within the range between 11.9 and 12.3 eV. 
Pulsed laser radiation in a wide wavelength range from $237$ to $289$~nm is found to provide efficient photodissociation of molecular ions that are formed in reactions of Th$^+$ with impurities in the buffer gas, leading to a significantly increased storage time for Th$^+$ in the ion trap.
\end{abstract}

\pacs{32.30.Jc 		
     32.80.Rm 		
     42.62.Fi,    
     37.10.Ty     
      }   

\maketitle

\section{Introduction}\label{sec:intro}

While the energy scales of radiative transitions in the atomic electron shell and within the nucleus are usually separated by several orders of magnitude, the nuclear transition of $^{229}$Th at about 8~eV \cite{Beck:2007, Beck:2009} offers a rare  opportunity to study the regime of near-degeneracy between both types of excitations. In this case, the coupling between the electronic and nuclear moments through hyperfine interaction may lead to a strong enhancement of radiative nuclear excitation and decay rates in so called electronic bridge processes \cite{Krutov:1968, Morita:1973, Crasemann:1973, Matinyan:1998, Tkalya:2003}.
Taking into account the known properties of the low-energy nuclear transition in $^{229}$Th -- a magnetic dipole transition at $7.8(5)$~eV -- 
one may consider different atomic configurations for an experimental study of these effects. In neutral Th, the ionization potential of 6.3~eV \cite{Goncharov:2006} lies below the nuclear excitation energy, so that photoionization or internal conversion compete with electronic bridge excitation or decay, respectively. This is not the case for positive thorium ions of all charge states, because of their higher ionization potentials. With a complex electronic level structure resulting from   three valence electrons and with an ionization potential of about 12~eV, Th$^+$ seems to offer the highest probability of finding suitable electronic transitions close to the nuclear resonance. Alternative options may be considered in Th$^{3+}$ or Th$^{89+}$. Th$^{3+}$ possesses a single valence electron and the level density around 8~eV is low, but, it permits laser cooling and sensitive high-resolution spectroscopy \cite{Campbell}. In hydrogen-like Th$^{89+}$ the strong hyperfine coupling of the spins of the nucleus and the $1s$ electron is expected to lead to a big enhancement of the nuclear radiative transition rate \cite{Karpeshin}.

A comprehensive analysis of the 13 lowest three-electron configurations of Th$^+$, consisting of 5f, 6d, 7s and 7p electrons has identified 461 of the 497 levels predicted for these configurations~\cite{Zalubas:1974}. 
Experimental energy values are published for 271 levels of even parity up to $\un{7.4}{eV}$ excitation energy and for 236 levels of odd parity up to $\un{8.1}{eV}$ excitation energy~\cite{Wyart:web}. Taking higher electron configurations into account, one may expect a high number of yet unidentified levels above 7~eV and in the vicinity of the $^{229}$Th nuclear transition energy. 
Based on relativistic Hartree-Fock calculations it has been shown that the energy level density in Th and Th$^+$ increases approximately exponentially with the excitation energy \cite{Dzuba:2010}. No systematic experimental search for levels of Th$^+$ that are suitable for electronic bridge excitation has been reported so far.  

Here we present the results of an experiment with two-step laser excitation of trapped $^{232}$Th$^+$ ions, covering the excitation energies between 7.3 and 8.3~eV, i.e. the $1\sigma$-search range for the nuclear excitation according to Refs.   \cite{Beck:2007, Beck:2009}. Previously, laser excitation of Th$^+$ has been reported on a few transitions at lower excitation energy \cite{Kaelber:1989, Herrera:2012}. In our experiment,
the first excitation step is done on the resonance line at 402~nm wavelength which couples the (6\emph{d}$^{2}$7\emph{s})\emph{J}=3/2 ground state with the (6\emph{d}7\emph{s}7\emph{p})\emph{J}=5/2 state at 24874~cm$^{-1}$. In order to obtain information on the angular momentum of the excited states, the line at 399.6~nm to the (5\emph{f}6\emph{d}$^2$)\emph{J}=1/2 state at 25027~cm$^{-1}$ is used alternatively. In the following we label states by their energy in cm$^{-1}$ and their total angular momentum $J$ (see Fig. 1). The second excitation step requires a tunable laser that covers the wavelength range between 237 and 294 nm. For this, we employ the third harmonic radiation of a pulsed titanium sapphire (Ti:Sa) laser. In future experiments, the same laser systems can be used to search for the two-photon electronic bridge excitation of $^{229}$Th$^+$ \cite{PorsevPeik:2010}.

\begin{figure}[b!th]
\centering
\includegraphics[width=0.35\textwidth]{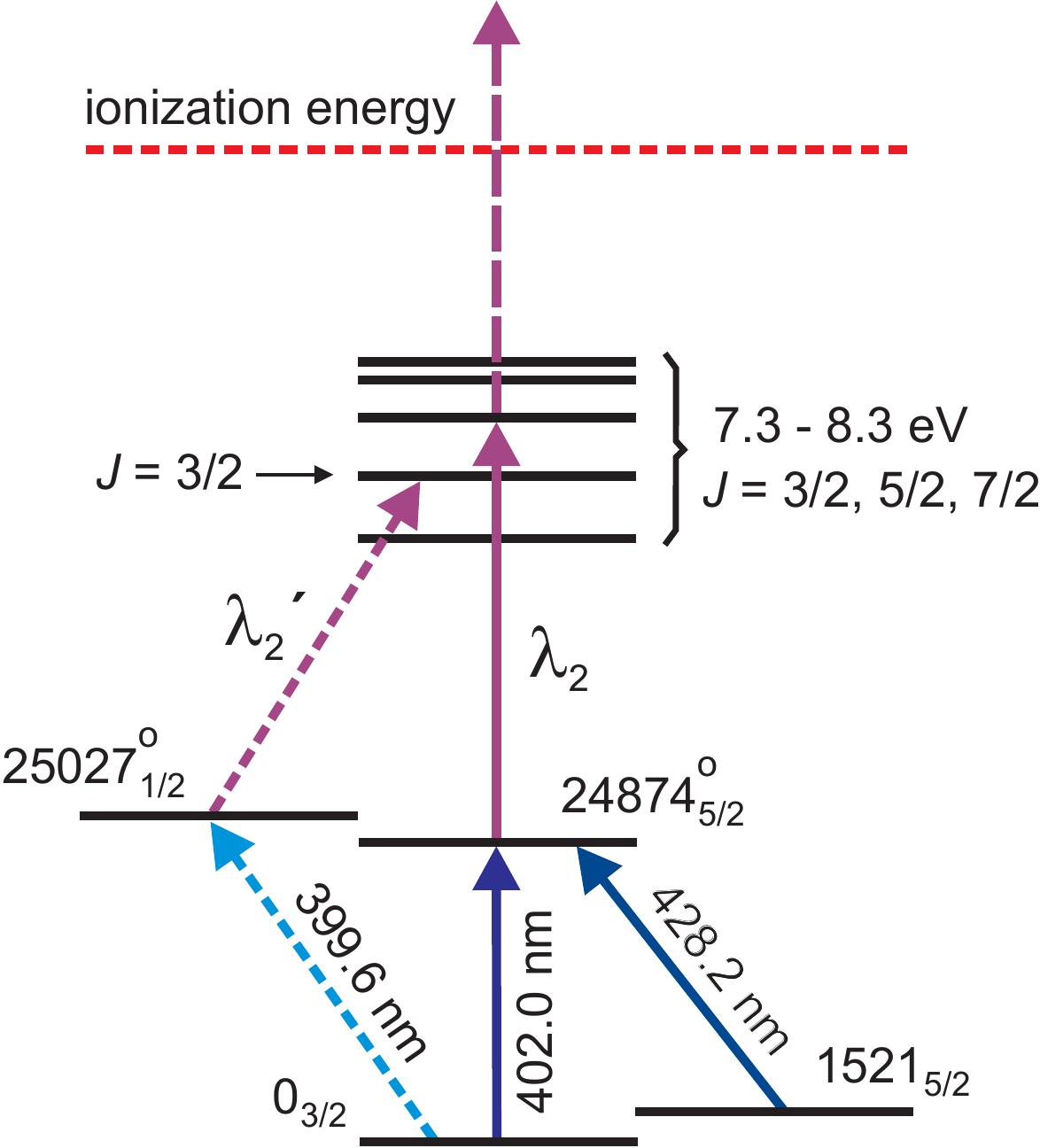}
\caption{\label{fig:level}
{Partial level scheme of Th$^+$, showing the transitions relevant to our investigations. The transition at 402~nm serves as the first excitation step.   $\lambda_{2}$ and $\lambda^{'}_{2}$  correspond to wavelengths of the second excitation step produced by a pulsed Ti:Sa laser. Excitation at 399.6~nm is used for the identification of high-lying states with $J=3/2$.
The level energies are given in $\mathrm{cm}^{-1}$ and the subscript denotes the total angular momentum $J$. A superscript (o) indicates odd parity.}
}
\end{figure}

\section{Experimental}\label{sec:experimental}

We use a segmented linear Paul trap, described in~\cite{Herrera:2012}, to trap a cloud of around $10^6$ buffer-gas cooled $^{232}$Th$^+$ ions. A schematic of the experimental setup is shown in Fig. 2. The trap is loaded by ablating a metallic thorium target using a Nd:YAG laser emitting $\un{5}{ns}$ pulses with an energy of $\leq\un{1}{mJ}$. 
The light for the first excitation step is provided by an extended-cavity diode laser (ECDL) with a maximum output power of $\un{7}{mW}$ at $\un{402}{nm}$, shaped into pulses of typically $\un{50}{ns}$ length by a fast acousto-optical modulator (AOM). The decay of the intermediate $24874_{5/2}$ state populates several metastable levels. We actively deplete the lowest of these levels ($1521_{5/2}$) with a frequency doubled ECDL continuously emitting light at $\un{428}{nm}$. By using argon at $\un{0.2}{Pa}$ pressure as the buffer gas, the remaining levels are collisionally quenched with sufficient efficiency~\cite{Herrera:2012}.

The light for the second excitation step is produced by a pulsed nanosecond Ti:Sa laser (Photonics Industries model TU-L) that is synchronized with the pulses from the ECDL. A system for single-pass third-harmonic generation (THG) provides a tuning range from $237$ to $\un{297}{nm}$. After mode cleaning and shaping the beam to a diameter of $\un{1}{mm}$, an average power of $\un{10}{mW}$ is typically available at the trap. At a repetition rate of $\un{1}{kHz}$ and a nominal pulse length of $\un{20}{ns}$ FWHM, this corresponds to $\approx\un{0.5}{kW}$ peak power. The emission spectrum typically consists of about 10 to 20 longitudinal modes separated by the free spectral range $\delta\nu_\mathrm{FSR}=\un{0.6}{GHz}$ of the laser resonator. The beams from the Ti:Sa and ECDL lasers are superposed and aligned along the axis of the ion trap. 

The fluorescence of the excited Th$^+$ ions is detected using photomultipliers (PMT) with different spectral response. One  photomultiplier (PMT1) serves to monitor the population in the intermediate $24874_{5/2}$ state and is equipped with a bandpass filter that transmits only wavelengths near 405~nm, within a 10 nm wide range. A second photomultiplier (PMT2) detects fluorescence resulting from the decay of higher energy levels excited by the second excitation step. 
It is sensitive to wavelengths between 300 and $\un{650}{nm}$, with a notch filter that blocks the $\un{402}{nm}$ fluorescence from the decay of the intermediate state. A third PMT provides sensitivity in the vacuum-ultraviolet range from 110~ to 200~nm.
Fast gated integrators are used to evaluate the PMT signals only during a limited detection window that begins after the end of the excitation pulses and has a duration of about  $\un{100}{ns}$. 
The signals are integrated over several hundred pulses and read out by the control computer.

\begin{figure}[h!tb]
\centering
\includegraphics[width=0.45\textwidth]{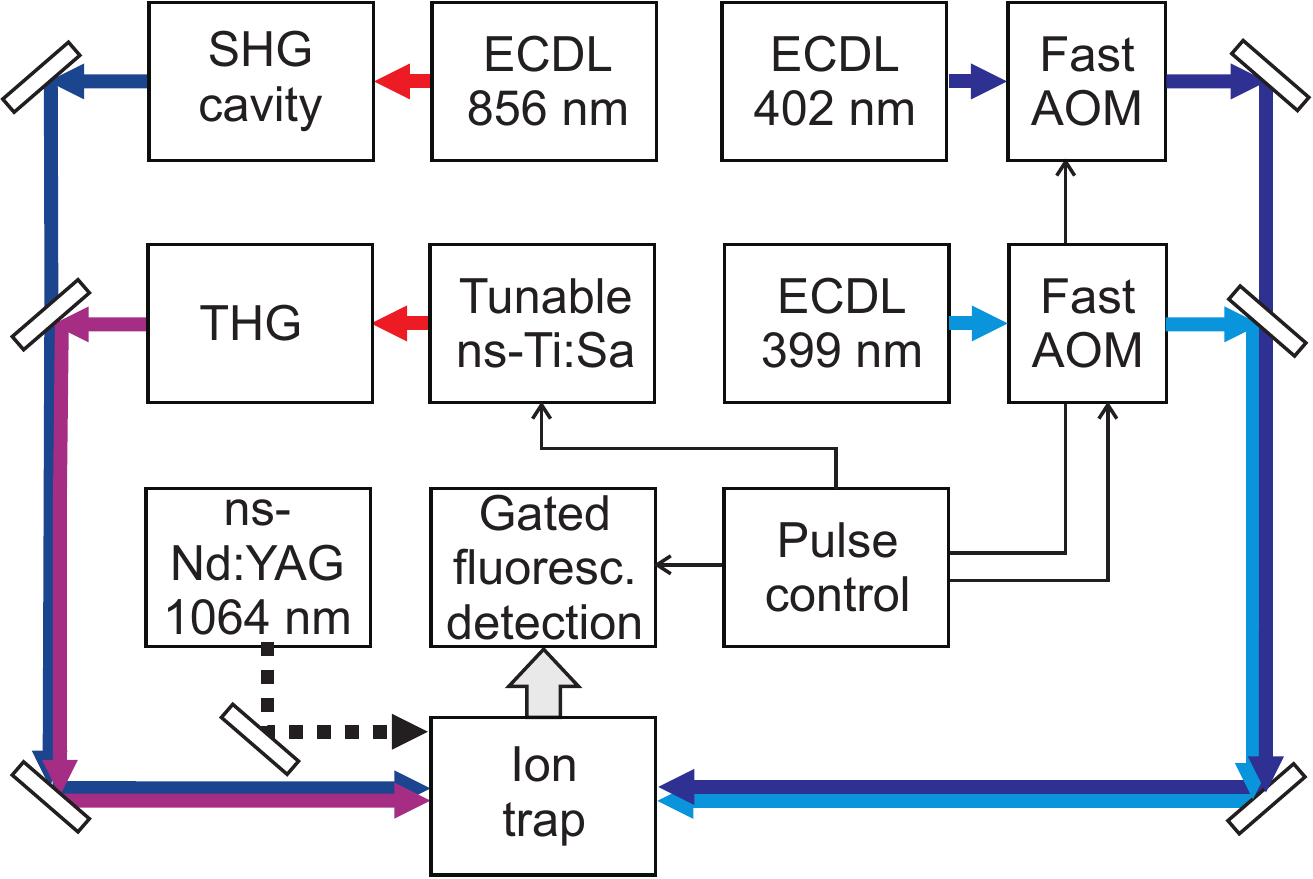}
\caption{\label{fig:optical} Block diagram of the experimental setup for laser excitation of trapped Th$^+$ ions.}
\end{figure}

\section{Newly observed states}\label{sec:two-photon}

\begin{figure}[b!th]
\centering
\includegraphics[width=0.45\textwidth]{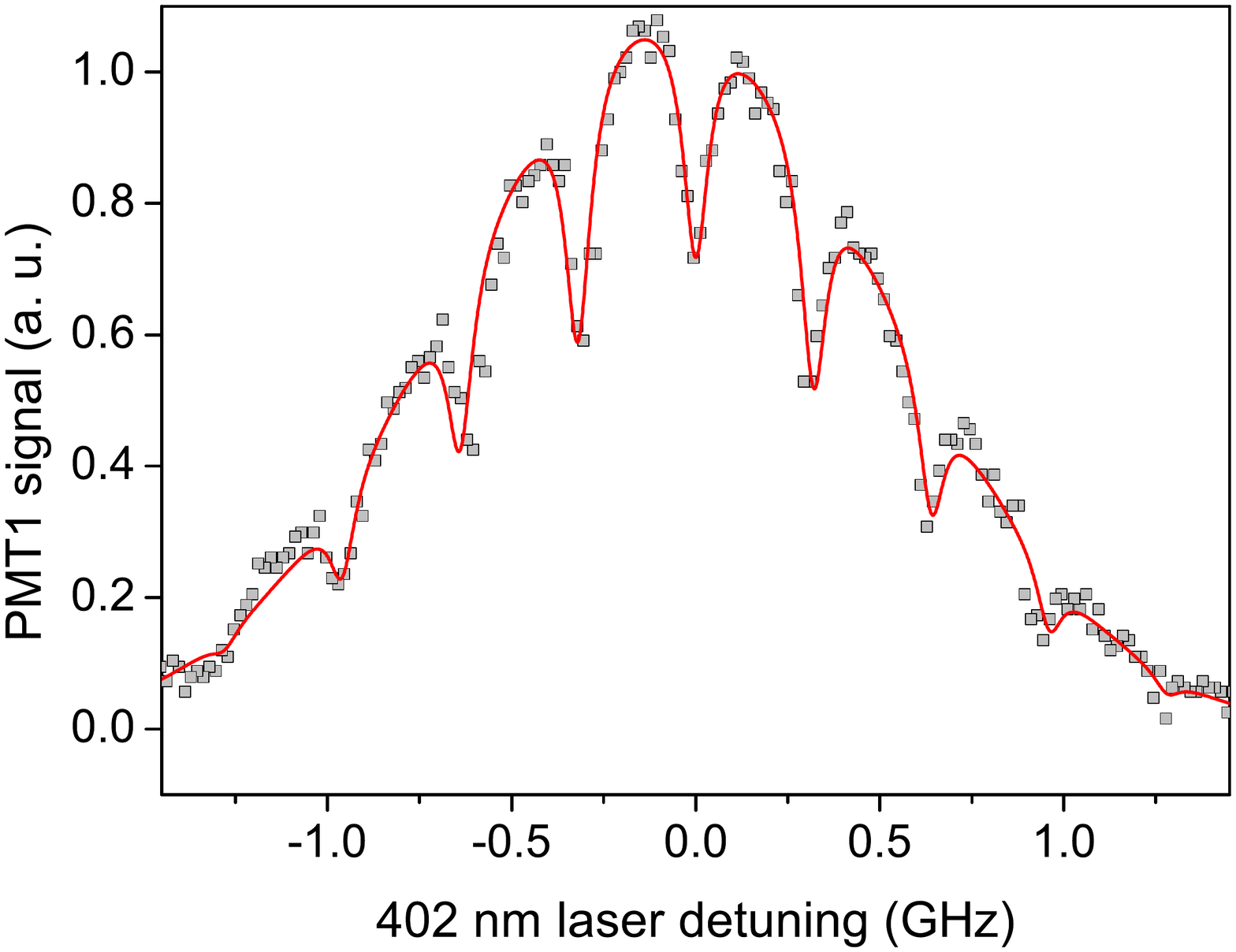}
\caption{\label{fig:pulseTPE3}
Laser excitation spectrum of the $\un{402}{nm}$ line of trapped Th$^+$ ions, showing multiple dips resulting from excitation of a higher energy level with several longitudinal modes of the Ti:Sa laser within the Doppler-broadened line profile.
The solid line shows a fit with a Gaussian lineshape with equidistant Lorentzian depletion resonances.}
\end{figure}

\begin{figure}[b!th]
\centering
\includegraphics[width=0.45\textwidth]{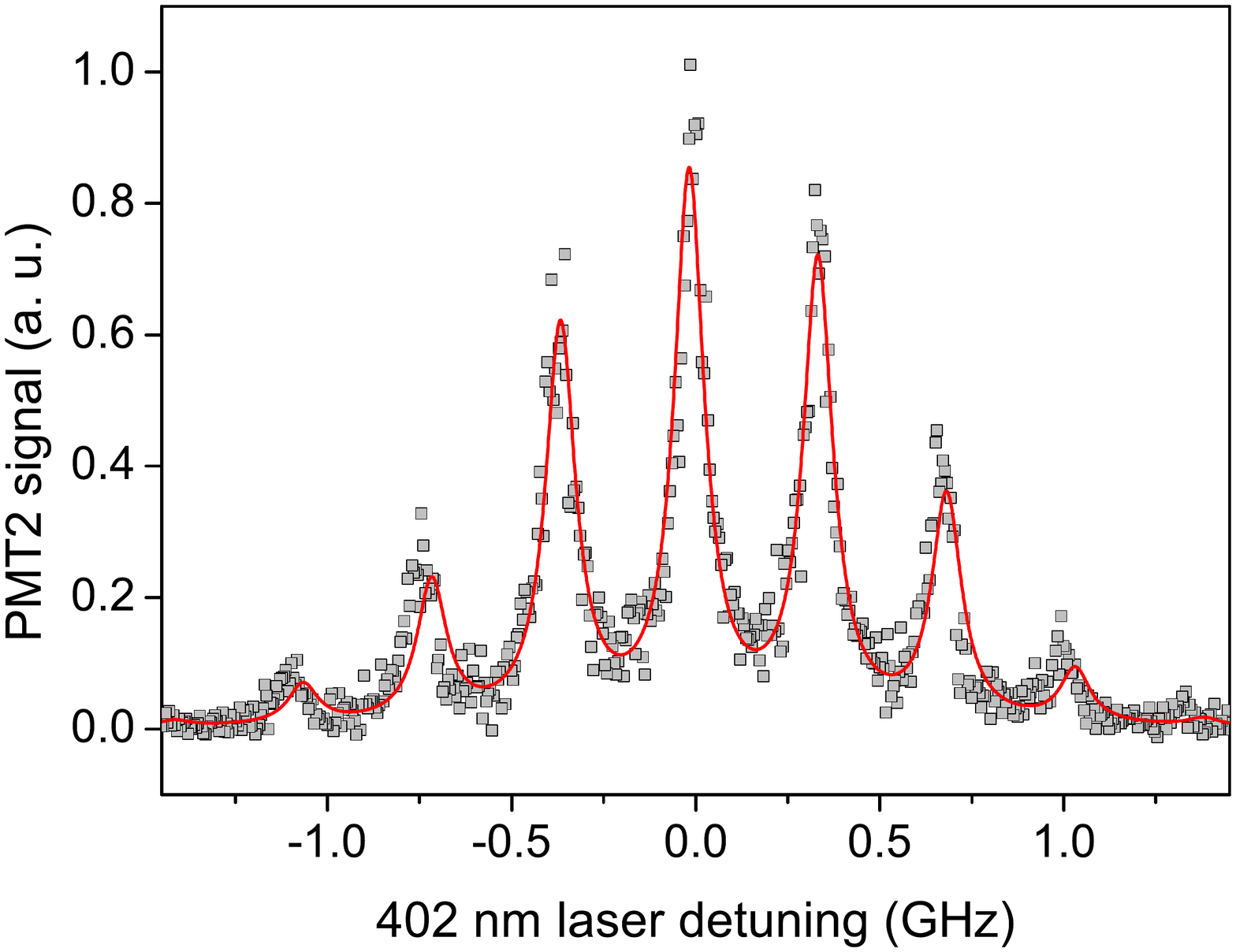}
\caption{\label{fig:pulseTPE4}
Fluorescence emission of trapped $^{232}$Th$^+$ ions resulting from two-photon laser excitation registered with a broadband photomultiplier (300 to 650~nm). The peaks arise from excitation with different longitudinal modes of the Ti:Sa laser. The solid line is a fit to equidistant Lorentzian lines under a Gaussian envelope.}
\end{figure}

In order to observe excited states in resonant two-photon excitation, we tune the Ti:Sa laser frequency by adjusting the intracavity diffraction grating with a stepper motor and a piezoelectric transducer for fine tuning. At the same time, the ECDL is scanned over the Doppler-broadened line of the $\un{402}{nm}$ transition. In Figs.~\ref{fig:pulseTPE3} and \ref{fig:pulseTPE4} we give example spectra where the Ti:Sa laser is held such that the frequencies of several longitudinal modes can excite transitions from the intermediate $24874_{5/2}$ level to a highly excited state.
We observe fluorescence signals from the high-lying states dominantly with the PMT that is sensitive in the visible and near-UV region, because the dense electronic level structure favors a decay through several intermediate states. Due to the short lifetimes of the states involved, the emission of fluorescence occurs mostly within a decay time of about 100~ns after the excitation pulses.

Figure~\ref{fig:pulseTPE3} presents the fluorescence signal detected with PMT1 around 402~nm as a function of the detuning of the ECDL driving the first excitation step. The Doppler-broadened spectrum shows 
multiple dips which demonstrate the depletion of the intermediate level population for the velocity classes that are resonantly excited to a high-lying state by longitudinal modes of the Ti:Sa laser. The Ti:Sa laser provides sufficient power to saturate almost all observed transitions and the resonant velocity classes show a significant depletion of the intermediate state population. First and second excitation steps have different excitation wavelengths $\lambda_1$ and $\lambda_2$ with different resulting Doppler sensitivities $k v$, where $k$ is the wave vector and $v$ is the velocity of interacting ions.  When scanning the ECDL, the depletion dips occur at a frequency spacing given by $\delta\nu = \left( \lambda_2 / \lambda_1 \right) \delta\nu_\mathrm{FSR}$. 

The main indicator for the two-photon excitation, however, is the fluorescence emission of the velocity groups transferred to the upper excited state by the Ti:Sa laser. Figure~\ref{fig:pulseTPE4} shows the fluorescence signal as a function of the ECDL detuning, detected with PMT2 where 402-nm light is blocked. For better resolution, the spectrum shown here is taken under conditions where the second excitation step is not saturated. 
Fluorescence detection in the vacuum-ultraviolet (VUV) was tested with some of the transitions and 
VUV emission was observed for excitation of the level at 60380.1~cm$^{-1}$. 
The predominant emission of photons in the visible and near-UV regions results from decay channels through several intermediate states. 

During the search for two-photon transitions, we have also observed about 150 single-photon transitions that are directly excited by the Ti:Sa laser. All these lines are among those listed in Ref~\cite{Zalubas:1974}. They originate either directly from the ground state (21 transitions) or from low-lying metastable levels (mainly from those with energies below 5000 cm$^{-1}$) that are populated by spontaneous decay from the $24874_{5/2}$ level or through buffer-gas collisions. They do not exhibit the characteristic spectra shown in figures~\ref{fig:pulseTPE3} and ~\ref{fig:pulseTPE4} and are therefore easily distinguished.

\begin{table}[b!th]
\centering
\caption{\label{table:levels}Total excitation energy for the observed energy levels, together with total angular momentum $J$ (where known), and parameter $S_f$ characterizing the pulse intensity required to saturate the fluorescence signal. With the possible exception of four cases marked with an asterisk, all levels have even parity and angular momenta of $3/2$, $5/2$ or $7/2$.}
\setlength{\tabcolsep}{3pt}
\begin{tabular}{lcr||lcr}
\hline
\hline
Level				& $J$ & $S_f$       & Level              & $J$ &   $S_f$      \\
(cm$^{-1}$)	&     & (kW/cm$^{2}$)  &  (cm$^{-1}$)    &     &   (kW/cm$^{2}$) \\
\hline
58875.5 		&	    & 200            & 64150.3     & 3/2 & 2.8   \\
59387.1     &     & 150            & 64560.4     & 3/2 & 7.0  \\
59477.4     & 3/2 & 0.62           & 64813.7     &     & 0.32  \\ 
59803.0     &     & 0.28           & 64860.4$^*$ &     & >450  \\
60380.1     & 3/2 & 2.2            & 64920.1     &     & 21  \\
60618.6     & 3/2 & 11          	 & 65037.7     & 3/2 & 6.5   \\
60721.3	    &     & 2.0            & 65144.4     &     & 1.1   \\
61032.4     & 3/2 & 0.34           & 65191.1$^*$ &     & >330  \\  
61388.0 		&     & 0.1            & 65730.4     & 3/2 & 0.55  \\
61428.6     &     & 0.54		       & 65738.1     &     & 44  \\
61726.3     & 3/2 & 4.0            & 65799.6     & 3/2 & 2.9  \\
61963.6		  & 3/2 &	0.36           & 65910.0     &     & 9.8   \\ 
62307.2     &     &	5.6		         & 65946.9     &     & 1.3   \\	
62373.8		  & 3/2 & 63             & 66052.0     &     & 20  \\  
62477.0     &     & 8.9            & 66141.2$^*$ &     & >110  \\  
62560.1     &     &	2.7		         & 66333.7     & 3/2 & 16  \\
62562.2     & 3/2 & 5.6		         & 66558.0     &     & 0.3   \\  
62753.1     &     &	7.7		         & 66609.0     &     & 4.2   \\	
63257.5     &     & 0.66           & 66702.9     & 3/2 & 64  \\
63298.4     &     & 27             & 66831.1		 & 3/2 & 0.28  \\
63557.7     &     & 19             & 66855.6     &     & 15	 \\
64122.0     &     & 10             & 67066.2$^*$ &     & >22   \\
\hline
\hline
\end{tabular}
\end{table}

While scanning the Ti:Sa laser over the range from 237 to $\un{294}{nm}$, two-photon excitations to 44 energy levels in the investigated energy range of 7.30 to $\un{8.31}{eV}$ were observed (see table~\ref{table:levels}). Only one of these levels was previously known, and its tabulated total energy value of $\un{59387.31}{cm}^{-1}$~\cite{Zalubas:1974} is in agreement with our measurement.
We estimate an uncertainty of $\un{0.2}{cm^{-1}}$ for the total energy, predominantly limited by the uncertainty of identifying the center of the Ti:Sa laser spectrum based on the reading of a Fizeau wavemeter. 

For each of the observed levels, we investigated the dependence of the fluorescence signal on the intensity of the Ti:Sa pulse. For almost all transitions the fluorescence saturates when high pulse intensities are applied, so that we can extract a saturation parameter $S_f$ by fitting with the expression $I_{Fl}\propto I/(I+S_f)$. 
In contrast to a two-level case, $S_f$ does not describe the overall population of the upper level, because this depends on the rates of different decay channels back to the ground state. 
All of the measured values of $S_f$ are within three orders of magnitude and within the range expected for electric dipole  transitions, taking into account the Doppler broadening, the laser spectral distribution and frequency modulation resulting from  micromotion of the ions in the trap \cite{Herrera:2012}. 
For four transtions we can only give a lower limit for $S_f$ because saturation was not observed with the available laser intensity.  

Assuming the observed transitions to have electric dipole character and taking into account the angular momentum $J'=5/2$ and odd parity of the intermediate state, selection rules require the levels in table~\ref{table:levels} to have even parity and angular momenta of $3/2$, $5/2$ or $7/2$. Four possible exceptions are the levels marked with an asterisk where saturation was not measured and where we can not exclude magnetic dipole or electric quadrupole character of the second excitation step. 

For each level, we have tested the alternative excitation path through the intermediate level $25027_{1/2}$ (see Fig.~\ref{fig:level}) by using the ECDL with a wavelength of $\un{399.6}{nm}$. 
In this way, levels that are excited via both intermediate states with $J=5/2$ and $J=1/2$ are identified as having an angular momentum $J=3/2$ and have been marked accordingly in table I.
For all these transitions there is no uncertainty in the electric dipole assignment based on the observed 
saturation intensities. 

The density of observed levels agrees with earlier ab-initio calculations~\cite{Porsev1+:2010}, which predicted 9 levels with $J=3/2$ in the energy range from $\un{58876}{cm^{-1}}$ to $\un{63955}{cm^{-1}}$, where we observed 8 such levels. However, since the calculations give energies only to within an uncertainty of about 10\% and because the experiment does not provide complete information on the electronic configuration, a direct comparison between observed and calculated levels is not possible.


\section{Level spacing statistics}

At the excitation energies investigated here, a strong mixing of different three-electron configurations of Th$^+$ is expected. In this case the well known approximative coupling schemes for the single-electron angular momenta are not applicable, and parity and total angular momentum $J$ are the only quantum numbers that characterize the energy levels. An empirical indication on the validity of this statement can be obtained from the distribution of adjacent level spacings, because in complex atomic and nuclear spectra a repulsion of energy levels characterized by the same parity and same value of $J$ is observed \cite{Rosenzweig:1960}. This level repulsion gives rise to a Wigner distribution of level spacings. In contrast, an ensemble of uncorrelated levels with different values of $J$ or another ''hidden'' angular momentum quantum number would show a Poisson distribution of spacings. 

In order to perform the analysis, we calculate the histogram of spacings $s$ between adjacent levels, normalized to the average observed level spacing and fit it with the Brody distribution~\cite{Brody:1981}
\begin{eqnarray}
P_q\left(s\right) 
= \left(1+q \right) \alpha \, s^q \, 
  \mathrm{exp}\left( -\alpha \, s^{q+1} \right)
\\
  \qquad \mathrm{where}\quad \alpha=\Gamma\left(\frac{2+q}{1+q}\right) \quad,
\nonumber
\end{eqnarray}
and $\Gamma(x)$ is the Gamma function.
\begin{figure}[bt]
\centering
\includegraphics[width=0.40\textwidth]{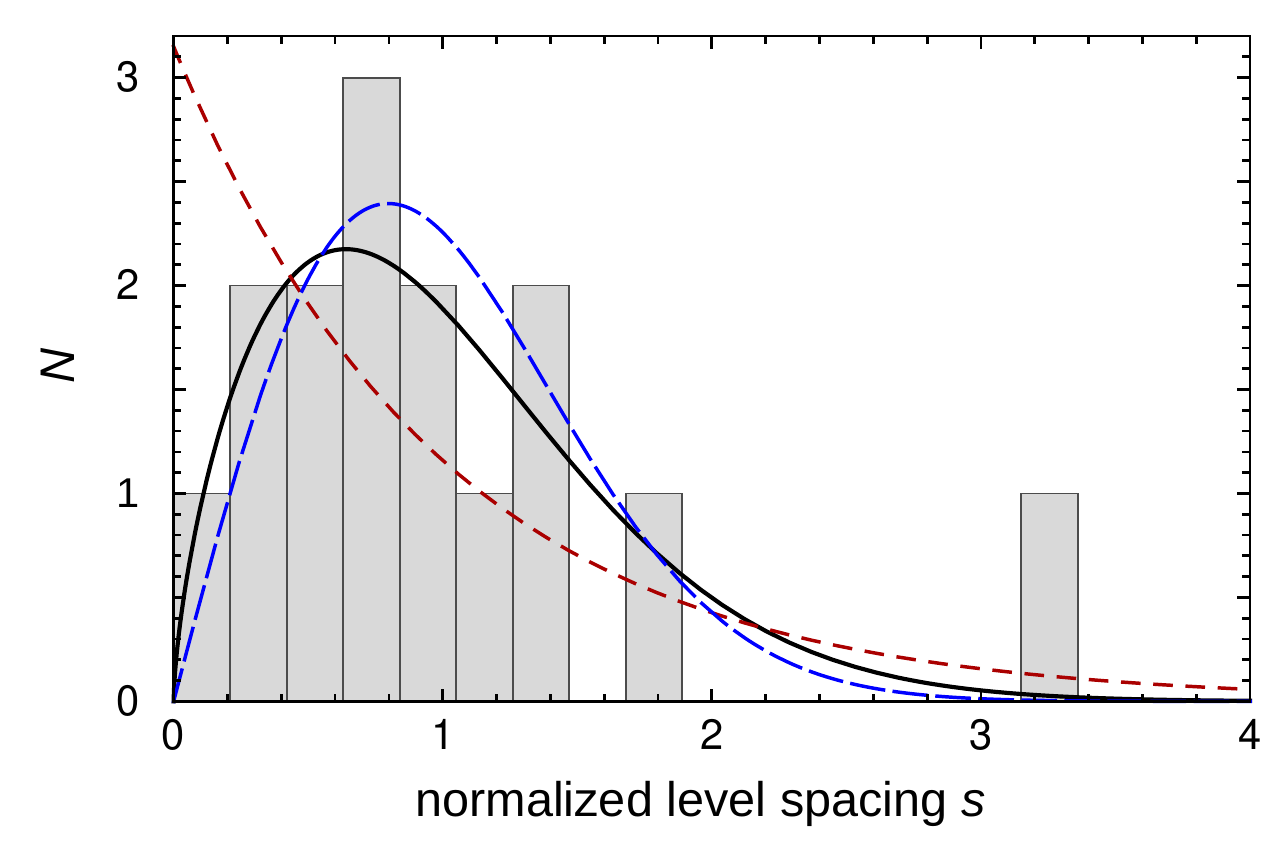}
\caption{\label{fig:histogram}
Histogram of the normalized adjacent level spacings $s$ for the set of observed levels with $J=3/2$. The thick black curve is a fit with the Brody distribution described in the text. The (blue) long-dashed line represents a fit to a Wigner, and the (red) short-dashed line to a Poisson distribution.
}
\end{figure}
For the parameter $q=0$, the Brody distribution takes on the form of the Poisson distribution expected for uniformly random distributed, non-interacting levels. For $q=1$ it becomes identical to the Wigner distribution expected when the level  repulsion is strong. 
Analyzing the set of newly observed $J=3/2$ states, we find significant repulsion indicated by $q=0.65$ (see Fig. \ref{fig:histogram}). The extracted value of $q$ depends somewhat on the binning of the histogram, but the variation is limited with a standard deviation of $0.1$. To test the statistical significance of this result, we have performed a Monte-Carlo simulation for a Poissonian level distribution. Even for the low available number of 15 spacings, a Brody  parameter $q=0.65$ was found to appear randomly with less than 5\% probability. The trend towards a Wigner distribution of level spacings at high excitation energies in Th$^+$ is also supported by the previously available data \cite{Wyart:web}. We perform the same analysis on subsets of 16 adjacent even-parity levels with $J=3/2$ over the range from $0$ to $\un{56235}{cm^{-1}}$ and find the Brody parameter growing from $q=0.1$ near the ground state to $q=0.6$ at the highest energies.
This indicates the increase of configuration mixing with growing excitation energy in the level scheme of Th$^+$. Furthermore, it may lead to the appearance of quantum chaotic behaviour, as has been theoretically described for the case of cerium, the lanthanide neighbor element to thorium in the f-block of the periodic table  
\cite{Flambaum:1999}.

\section{Multi-photon ionization in trapped T$\lowercase{\text{h}}$$^{+}$ ions}\label{sec:ionization}

\begin{figure}[bt]
\centering
\includegraphics[width=0.44\textwidth]{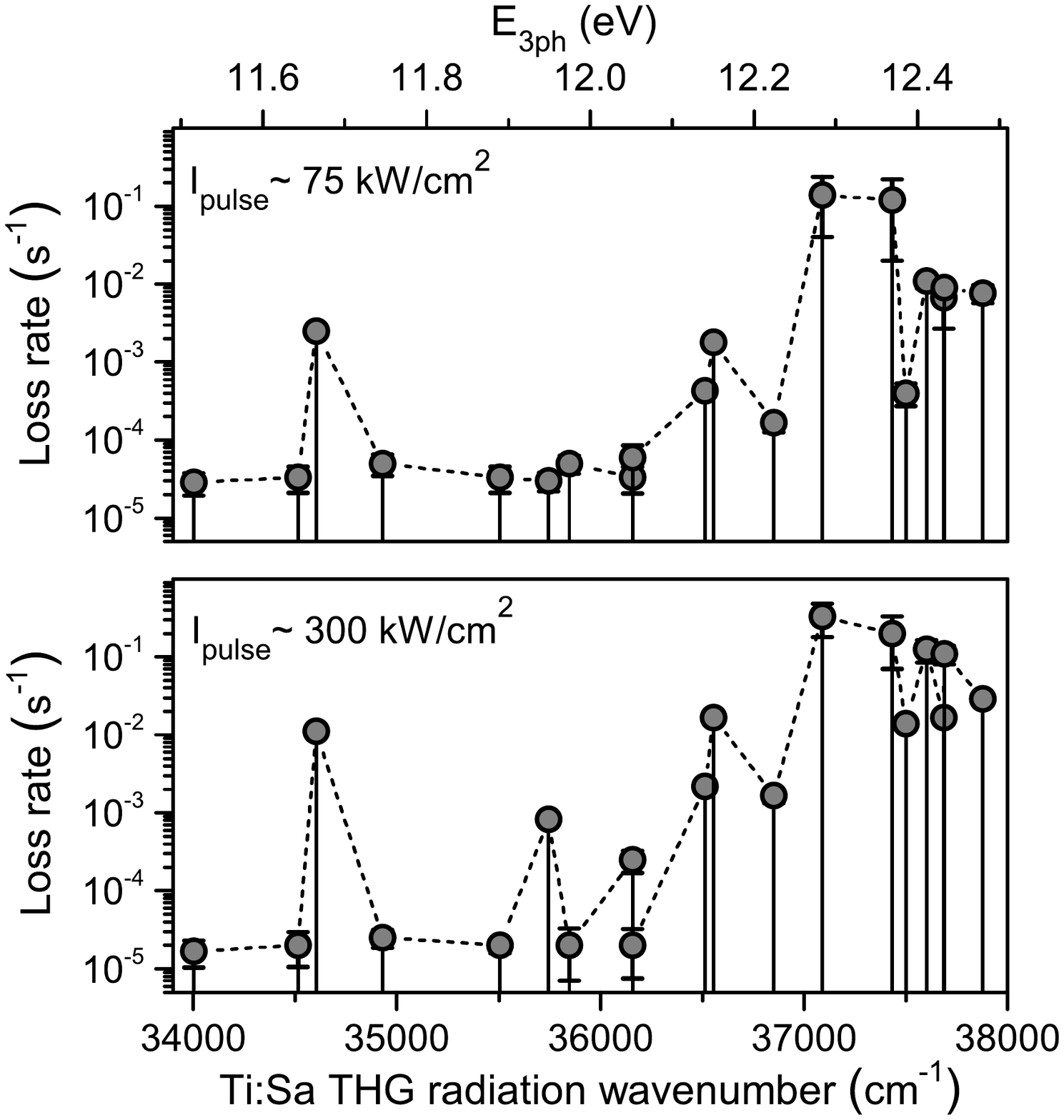}
\caption{\label{fig:ionization}
Th$^+$ ion loss rate as a function of the Ti:Sa third harmonic wavenumber for two values of the pulse intensity, indicating multi-photon ionization.     
For each data point the Ti:Sa radiation resonantly excites one of the levels listed in Table I. The levels can be identified by adding the intermediate state wavenumber 24874~cm$^{-1}$ to the value in the abscissa. 
The upper energy scale $E_{3ph}$ is the sum of the energies of one photon at 24874~cm$^{-1}$ and two photons from the Ti:Sa laser.}
\end{figure}

When the ions are irradiated with resonant pulses from the 402~nm ECDL and the Ti:Sa laser 
with intensities above $S_f$ (see Table I) and  radiation wavelengths shorter than 270~nm, we observe a fast exponential decay of the Th$^+$ fluorescence signals. The time constant of this decay is much shorter  than the typical storage  time ($\approx 3\times10^4$~s) observed in our trap in the absence of two-step excitation (see Section VI). 
We investigated the origin of the decay by ramping down the amplitude of the trap rf voltage and detecting the release of ions with a channeltron detector, so that information on the charge-to-mass ratio is obtained \cite{Zimmermann:2012}. 
Correlated with the decay of the Th$^+$ signals, 
we observed the appearance of a fraction of Th$^{2+}$. At the usual stationary operation conditions of our trap, both charge states are trapped stably. We attribute the formation of Th$^{2+}$ to a resonantly enhanced multi-photon ionization  (see Fig. 1), occuring via the absorption of further photons from the Ti:Sa radiation. 

Figure \ref{fig:ionization} shows measured loss rates of the trapped Th$^+$ ions  as a function of the wavenumber of the third harmonic of the Ti:Sa laser radiation, which is tuned to the observed levels (see Table I) for the second excitation step, while the 402 nm ECDL scans over the Doppler profile as described in Section 3.   The experiment is carried out for the levels between 58875.5 and 62753.1 cm$^{-1}$ which corresponds to third harmonic radiation wavelengths from 264 to 294 nm. Results are given for two different pulse intensities of the Ti:Sa laser. 

For Ti:Sa laser wavenumbers above 36500 cm$^{-1}$ we observe formation of Th$^{2+}$ ions with a rate above $10^{-4}$~s$^{-1}$ for all studied levels. This wavenumber range corresponds to excitation energies above 12.14~eV, 
calculated as the sum of the energies of one photon from the 402 nm ECDL and two UV photons from the Ti:Sa laser. In the following, this quantity is denoted as $E_{3ph}$. 

In the region $E_{3ph}< 12.05$~eV no formation of Th$^{2+}$ ions is observed for the lower intensity for most of the measurements. An exception is the transition to the state at 59477.4 cm$^{-1}$ (corresponding to $E_{3ph} = 11.67$~eV). 
This  can be explained with a four-photon process where 
Th$^+$ is excited from the level 59477.4 cm$^{-1}$ to a bound state close  to the ionization threshold and is then ionized to Th$^{2+}$ through absorption of an additional photon. For the state  60618.6 cm$^{-1}$ ($E_{3ph} =11.95$~eV) ionization appears only for high pulse intensity. The excitation of the state 61032.4 cm$^{-1}$ ($E_{3ph} =12.05$~eV) does not lead to ionization but Th$^{2+}$ ions appear when the UV radiation of the Ti:Sa laser is detuned out of resonance by $-1.5$~cm$^{-1}$. For these two transitions an ionization through a four-photon process may also be assumed.

Interpreting the rapid loss of Th$^+$ for wavenumbers above 36500 cm$^{-1}$ as the onset of three-photon ionization, one can deduce an ionization potential of Th$^+$ slightly above 12.0~eV. The possible mixture of three- and four-photon ionization, however, does not allow us to observe a clear threshold in the ionization rate.      
Because of this complexity we infer from the data presented in Fig. \ref{fig:ionization} that the ionization potential lies  in the range between 11.9 and 12.3 eV.  A more precise experimental determination would be feasible by employing an independently tunable laser for the third excitation step close to the ionization threshold. 

Ab initio calculations of the second ionization potential of thorium obtained values in the range between 11.08 and 12.87 eV \cite{Harding, Liu, Cao}. To the best of our knowledge no experimental measurement of this quantity has been published before. Two values sometimes cited as ''experimental'' are actually the results of extrapolations based on spectral data from other elements: $11.5(1.0)$~eV \cite{Finkelnburg} and  $11.9(1)$~eV \cite{Sugar}, and are in agreement with our observations. 

\section{Formation and photodissociation of molecular ions with T$\lowercase{\text{h}}^{+}$}
\label{sec:Photodissociation}

Previous studies have shown that Th$^+$ ions in the gas phase are highly reactive with compounds of oxygen (O$_2$, H$_2$O, NO, CO$_2$, etc.) and hydrocarbons \cite{Johnsen:1974, Joaquim:1996, Cornehl:1997}. In the presence of oxygen, the ThO$^+$ molecule is formed, characterized by a  double bond of 9.1~eV dissociation energy  \cite{Goncharov:2006}.
The remaining single valence electron gives rise to further reactions leading to   ThOH$^+$, ThO$_2^+$ and  ThO$_2$H$^+$ \cite{Cornehl:1997,Zhou:2010}. In the presence of CH$_4$, Th$^+$ will react to  ThCH$_2^+$, possessing a dissociation energy of 4.8~eV \cite{Joaquim:1996}.
In our experiment, ultra-high vacuum conditions with a base pressure in the $10^{-8}$~Pa range are ensured, with water and hydrocarbons as the dominant residual gases, and highly purified buffer gas (Ar with less than $1\times 10^{-6}$ impurity content) is used. Nevertheless, the fluorescence signals from Th$^+$ are only observable for a few hundred seconds~\cite{Herrera:2012} due to the formation of molecular compounds. Similar observations have been made in other experiments with trapped Th$^+$ and Th$^{3+}$ ions  \cite{Kaelber:1992,Churchill:2011}. 

When the trapped ions are exposed to Ti:Sa laser radiation with an intensity above $\un{50}{kW/cm^2}$  the observation time for Th$^+$ fluorescence is greatly extended to values of more than $3\times\un{10^4}{s}$.  The effect was observable for practically all laser wavelengths between  237 and 289~nm, i.e. nearly for the full wavelength range studied here during the search for two-step excitation of Th$^+$. 
It is in fact possible to let the fluorescence decay to near zero with the Ti:Sa laser switched off over a time of 30 minutes and then recover nearly the full original signal by turning the laser on. This can be attributed to photodissociation of trapped molecular Th-compounds.
At lower laser intensity a wavelength dependence of the photodissociation rate appears, but this was not the subject of further study. From the binding energies given above, it can be seen that photodissociation of ThCH$_2^+$ and $\mathrm{ThO}^+$, can only be obtained as a two- or three-photon process, respectively, in this wavelength range.
Because the molecular ions remain trapped, however, even a photodissociation rate as low as $10^{-2}$~s$^{-1}$ is sufficient to maintain a significant fraction of Th$^+$ ions in the trap. 

To gain information on the molecular species that are formed, we applied an additional alternating voltage of tunable frequency to two electrodes of the central segment of the trap. This ejects ions with specific mass-to-charge ratios from the trap when the applied frequency is equal to the ions' secular motional frequency. Molecular ions are allowed to form by turning off the Ti:Sa laser. Certain excitation frequencies prevent the recovery of the fluorescence signal when the Ti:Sa is switched on again. These correspond to masses between 246 and 259~amu, including $\mathrm{ThCH_2}^+$, $\mathrm{ThO}^+$, and $\mathrm{ThOH}^+$, all of which are likely products of reactions with impurities in the buffer gas. Applying the secular frequency corresponding to a mass of 265~amu does not suppress the fluorescence recovery, indicating the absence of $\mathrm{ThO}_2^+$.

Given that the formation of several reaction products is expected \cite{Johnsen:1974, Joaquim:1996, Cornehl:1997}, it is remarkable that photodissociation with a single laser leads to the establishment of an essentially stable Th$^+$ fraction. It seems that no stable molecular ion is produced here that can not be dissociated with the laser. 
This observation is of great practical importance for the planned experiments  with the radioactive $^{229}$Th$^+$ ions where the amount of substance available for loading the trap will be limited.

\section{Prospects for electronic bridge excitation}

Finally, we discuss an order of magnitude estimate of an electronic bridge excitation rate that can be expected with the experimental setup presented here. It is assumed that radiation of spectral width $\Delta\nu_{l}$ drives an electronic transition of width $\Delta\nu_{el}$, that is significantly larger than the linewidth of the isolated nuclear transition $\Delta\nu_{nu}$. $\Delta\nu_{n,e}$ denotes the frequency difference between electronic and nuclear resonance. Under the assumption that $\Delta\nu_{n,e} \gg \Delta\nu_{el}\gg\Delta\nu_{nu}$, the electronic bridge excitation rate is enhanced in comparison to a direct excitation of the nucleus by a factor $K\approx (E_{M1}^2 \Delta\nu_{el})/(\Delta\nu_{n,e}^2\Delta\nu_{nu})$, where $E_{M1}$ is the magnetic dipole coupling energy between electron and nucleus (in frequency units) \cite{tkalya1,Tkalya:2003}. The orders of magnitude for these parameters are given by $\Delta\nu_{nu}\sim 10^{-3}$~Hz, $\Delta\nu_{el}\sim 10^{7}$~Hz, $E_{M1}\sim 10^{9}$~Hz \cite{tkalya1} and $\Delta_{n,e}\sim 3\cdot 10^{12}$~Hz, according to the level density observed here. This leads to an enhancement in the range $K\sim 10^3$. In a calculation of the frequency-dependence of the enhancement factor $K$ for resonant excitation of the nuclear transition based on ab-initio atomic structure calculations, Porsev at al. \cite{PorsevPeik:2010} find a minimum value of $K\sim 10$ in a region far from strong electronic resonances and higher values elsewhere.     

To calculate the excitation probability $P=(c^2 \tau \delta_{nu} I)/(4h \nu^3 \Delta\nu_{l})$ we take the following parameters of our system:
pulse duration $\tau=20$~ns,  $\delta_{nu}=K \Delta\nu_{nu} \sim  1$~Hz (the effective linewidth of the nuclear resonance),
$\Delta\nu_{l}\approx 10^9$~Hz (determined by the inhomogenous linewidth of the ion cloud), 
 and $I\approx 10^5$~W/cm$^2$. This yields an excitation probability $P\sim 10^{-4}$ per laser pulse. In an experiment with a laser repetition rate of $10^3$~Hz and with more than$10^5$ ions, this should make the detection of the nuclear excitation feasible.

While scanning the Ti:Sa laser over the search range for the nuclear excitation the electronic resonances need to be taken into account. Single-photon excitation with the Ti:Sa laser from the ground state or a low-lying metastable level reduces the population of the intermediate $24874_{5/2}$ state and consequently the obtainable electronic bridge excitation rate. Another critical point is the resonantly enhanced three-photon ionization discussed in Sect. V because it leads to a loss of trapped $^{229}$Th$^+$. This should be avoided in the interest of using a radionuclide source of low activity. Because the multi-photon ionization has a nonlinear intensity dependence, it can be suppressed by reducing the Ti:Sa pulse power for wavelengths that are close to resonance with an electronic excited state. 
If the nuclear transition lies in the affected range of excitation energies, the small value of $\Delta\nu_{n,e}$ ensures 
that a high electronic bridge excitation rate can be obtained nevertheless.  
             
The position of the isomer energy relative to the electronic level structure will also determine the lifetime and radiative decay of the isomeric state. Three possible scenarios can be envisaged for the detection of the nuclear excitation. 
In the case of the isomer undergoing direct radiative decay, delayed VUV photons can be detected between the laser pulses. More likely, the strongly enhanced electronic bridge processes will dominate the decay of the isomer and will lead to the emission of characteristic photon cascades from the electron shell. The signal will be recorded with time resolved fluorescence detection in different spectral channels. Test experiments with $^{232}$Th$^+$ under identical laser parameters can be used to distinguish the electronic bridge nuclear decay from any purely electronic fluorescence process. Finally,
if the lifetime of the isomer is sufficiently long to allow population of $^{229m}$Th$^+$ to build up over several laser pulses, the most sensitive detection method will be high resolution spectroscopy of the 402~nm transition that will detect these ions based on their modified nuclear spin and hyperfine structure \cite{Peik:2003}. 

\begin{acknowledgments} We thank Chr. Tamm, A. Yamaguchi and K. Zimmermann for helpful discussions and contributions to the experimental setup and T. Leder for providing expert technical support. 
OAHS acknowledges support from DAAD (Grant No. A/08/94804), ITCR (Grant No. 83-2008-D), and MICIT (Grant No. CONICIT 079-2010).
\end{acknowledgments}


\begin{thebibliography}{95}

\bibitem{Beck:2007} B. R. Beck, J. A. Becker, P. Beiersdorfer, G. V. Brown, K. J. Moody, J. B. Wilhelmy, F. S. Porter, C. A. Kilbourne, R. L. Kelley, Phys. Rev. Lett. {\bf 98}, 142501 (2007).

\bibitem{Beck:2009} B. R. Beck, J. A. Becker, P. Beiersdorfer, G. V. Brown, K. J. Moody, C. Y. Wu, J. B. Wilhelmy, F. S. Porter, C. A. Kilbourne, R. L. Kelley, Lawrence Livermore National Laboratory, Internal Report {\bf LLNL-PROC-415170} (2009).

\bibitem{Krutov:1968} V. A. Krutov, V. N. Fomenko, Ann. Physik 7, {\bf 21}, 291 (1968).

\bibitem{Morita:1973} M. Morita, Prog. Theor. Phys. {\bf 49}, 1574 (1973).

\bibitem{Crasemann:1973} B. Crasemann, Nucl. Instrum. Meth. {\bf 112}, 33 (1973).

\bibitem{Matinyan:1998} S. Matinyan, Physics Reports {\bf 298}, 199 (1998).

\bibitem{Tkalya:2003} E. V. Tkalya, Physics-Uspekhi {\bf 46}, 315 (2003).

\bibitem{Goncharov:2006} V. Goncharov, M. C. Heaven, J. Chem. Phys. {\bf 124}, 064312 (2006).

\bibitem{Campbell} C. J. Campbell, A. G. Radnaev, A. Kuzmich, Phys. Rev. Lett. {\bf 106}, 223001 (2011).

\bibitem{Karpeshin} F. F. Karpeshin, S. Wycech, I. M. Band, M. B. Trzhaskovskaya, M. Pf\"utzner, J. Zylicz, Phys. Rev. C {\bf 57}, 3085 (1998).

\bibitem{Zalubas:1974} R. Zalubas, C. H. Corliss, J. Res. Natl. Bur. Stand. Sect. A {\bf 78A}, 163 (1974), [http://nistdigitalarchives.contentdm.oclc.org/cdm/ref/\\collection/p13011coll6/id/63425].

\bibitem{Wyart:web} J. Blaise, J.-F. Wyart, Selected constants, energy levels and atomic spectra of actinides,  [http://www.lac.u-psud.fr/Database/Contents.html].

\bibitem{Dzuba:2010} V. A. Dzuba, V. V. Flambaum, Phys. Rev. Lett. {\bf 104}, 213002 (2010).

\bibitem{Kaelber:1989} W. K\"alber, J. Rink, K. Bekk, W. Faubel, S. G¨oring, G. Meisel, H. Rebel, and R. C. Thompson, Z. Phys. A {\bf 334}, 103 (1989).

\bibitem{Herrera:2012} O. A. Herrera-Sancho,  M. V. Okhapkin, K. Zimmermann, Chr. Tamm, E. Peik, A. V. Taichenachev, V. I. Yudin, P. G\l{}owacki, Phys. Rev. A {\bf  85}, 033402 (2012).

\bibitem{PorsevPeik:2010} S. G. Porsev, V. V. Flambaum, E. Peik, Chr. Tamm, Phys. Rev. Lett. {\bf 105}, 182501 (2010).

\bibitem{Porsev1+:2010} S. G. Porsev, V. V. Flambaum, Phys. Rev. A {\bf 81}, 042516 (2010).

\bibitem{Rosenzweig:1960} N. Rosenzweig, C. E. Porter, Phys. Rev. {\bf 120}, 1698 (1960).

\bibitem{Brody:1981} T. A. Brody, J. Flores, J. B. French, P. A. Mello, A. Pandey, S. S. M. Wong, Rev. Mod. Phys. {\bf 53}, 385 (1981).

\bibitem{Flambaum:1999} V.V. Flambaum, A.A. Gribakina, G.F. Gribakin, I.V. Ponomarev, Physica D {\bf 131}, 205 (1999).

\bibitem{Zimmermann:2012} K. Zimmermann, M. V. Okhapkin, O. A. Herrera-Sancho, E. Peik, Appl. Phys. B \textbf{107}, 883 (2012).

\bibitem{Harding} J. H. Harding, P. J. D. Lindan, N. C. Pyper, J. Phys.: Condens. Matter {\bf  6}, 6485 (1994).

\bibitem{Liu} W. J. Liu, W. K{\"u}chle, M. Dolg,
Phys. Rev. A {\bf 58}, 1103 (1998).

\bibitem{Cao} X. Y. Cao, M. Dolg, Mol. Phys. {\bf 101}, 961 (2003).

\bibitem{Finkelnburg} W. Finkelnburg, W. Humbach, Naturwissenschaften {\bf 42}, 35 (1955). 

\bibitem{Sugar} J. Sugar, cited in: D. L. Hildenbrand, L. V. Gurvich, V. S. Yungman, The chemical thermodynamics of actinide elements and compounds, Part 13: The gaseous actinide ions, IAEA, Vienna, 1985.

\bibitem{Johnsen:1974} R. Johnsen, F. R. Castell, and M. A. Biondi, J. Chem. Phys. \textbf{61}, 5404 (1974).

\bibitem{Cornehl:1997} H. H. Cornehl, R. Wesendrup, M. Diefenbach, H. Schwarz, Chem. Eur. J. \textbf{3}, 1083 (1997).

\bibitem{Joaquim:1996} J. Mar\c{c}alo, J. P. Leal, A. P. de Matos, Int. J. Mass Spectrom. \textbf{157}, 265 (1996).

\bibitem{Zhou:2010} J. Zhou, H. B. Schlegel, J. Phys. Chem. A \textbf{114}, 8613 (2010).

\bibitem{Kaelber:1992} W. K\"alber, G. Meisel, J. Rink, R. C. Thompson, J. Mod. Opt. {\bf 39}, 335 (1992).

\bibitem{Churchill:2011} L. R. Churchill, M. V. DePalatis, M. S. Chapman, Phys. Rev. A \textbf{83}, 012710 (2011).

\bibitem{tkalya1} E. V. Tkalya, V. O. Varlamov, V. V. Lomonosov, S. A. Nikulin, Phys. Scr. {\bf 53}, 296 (1996).

\bibitem{Peik:2003} E. Peik, Chr. Tamm, Europhys. Lett. {\bf 61}, 181 (2003).

\end{thebibliography}
\end{document}